\documentclass[11pt]{article}
\def\iindex{{\em ...erased to make a blind copy...}\index}
\def\iindex{ }
\usepackage{amssymb}

\newtheorem{theorem}{Theorem}

\newtheorem{corollary}{Corollary}

\newtheorem{definition}{Definition}[section]
\newtheorem{remark}{Remark}

\newcounter{thanksnum}
\def\thanksnumber#1
{\setcounter{thanksnum}{\value{footnote}}\setcounter{footnote}{#1}%
                     \addtocounter{footnote}{-1}\footnotemark
                     \setcounter{footnote}{\value{thanksnum}}}
\def\newtheoremz#1{\@ifnextchar[{\@othmz{#1}}{\@nthmz{#1}}}

\def\@nthmz#1#2{%
\@ifnextchar[{\@xnthmz{#1}{#2}}{\@ynthmz{#1}{#2}}}

\def\@xnthmz#1#2[#3]{\expandafter\@ifdefinable\csname #1\endcsname
{\@definecounter{#1}\@addtoreset{#1}{#3}%
\expandafter\xdef\csname the#1\endcsname{\expandafter\noexpand
  \csname the#3\endcsname \@thmcountersepz \@thmcounterz{#1}}%
\global\@namedef{#1}{\@thmz{#1}{#2}}\global\@namedef{end#1}{\@endtheoremz}}}

\def\@ynthmz#1#2{\expandafter\@ifdefinable\csname #1\endcsname
{\@definecounter{#1}%
\expandafter\xdef\csname the#1\endcsname{\@thmcounterz{#1}}%
\global\@namedef{#1}{\@thm{#1}{#2}}\global\@namedef{end#1}{\@endtheoremz}}}

\def\@othmz#1[#2]#3{\expandafter\@ifdefinable\csname #1\endcsname
  {\global\@namedef{the#1}{\@nameuse{the#2}}%
\global\@namedef{#1}{\@thmz{#2}{#3}}%
\global\@namedef{end#1}{\@endtheoremz}}}

\def\@thmz#1#2{\refstepcounter
    {#1}\@ifnextchar[{\@ythmz{#1}{#2}}{\@xthmz{#1}{#2}}}

\def\@xthmz#1#2{\@begintheoremz{#2}{\csname the#1\endcsname}\ignorespaces}
\def\@ythmz#1#2[#3]{\@opargbegintheoremz{#2}{\csname
       the#1\endcsname}{#3}\ignorespaces}

\def\@thmcounterz#1{\noexpand\arabic{#1}}
\def\@thmcountersepz{.}
\def\@begintheoremz#1#2{ \trivlist \item[\hskip \labelsep{\bf #1\ #2}]}
\def\@opargbegintheoremz#1#2#3{ \trivlist
      \item[\hskip \labelsep{\bf #1\ #2\ (#3)}]}
\def\@endtheoremz{\endtrivlist}

\def\e{\varepsilon}

\def\defi{\stackrel{{\scriptscriptstyle \Delta}}{=}}

\def\d{\delta}
\def\o{\omega}

\def\F{{\cal F}}
\def\w{\widehat}

\def\R{{\bf R}}
\def\E{{\bf E}}
\def\P{{\bf P}}

\def\b{\beta}
\def\s{\delta}
\def\g{\gamma}
\def\C{{\bf C}}

\def\ww{\widetilde}

\def\t{\theta}
\def\oo{\bar}
\def\s{\sigma}

\def\A{{\cal A}}
\def\M{{\cal M}}

\newcommand{\be}{\begin{equation}}
\newcommand{\ee}{\end{equation}}
\newcommand{\bd}{\begin{displaymath}}
\newcommand{\ed}{\end{displaymath}}
\newcommand{\ba}{\begin{array}{ll}}
\newcommand{\ea}{\end{array}}
\newcommand{\baa}{\begin{eqnarray}}
\newcommand{\eaa}{\end{eqnarray}}
\newcommand{\baaa}{\begin{eqnarray*}}
\newcommand{\eaaa}{\end{eqnarray*}}   \font\sm=cmr10

\def\ww{\tilde}

\def\O{\Omega}

\date{Submitted: September 21, 2012. Revised: May 30, 2013 }
\title{On statistical indistinguishability of the complete and incomplete markets}
\author{
Nikolai Dokuchaev\\
 {\sm Department of Mathematics \& Statistics, Curtin
University,}\\ {\sm  GPO Box U1987, Perth, 6845 Western Australia} }
\begin{document}
\maketitle
\begin{abstract}
The possibility of statistical evaluation of the market completeness
and incompleteness is investigated for continuous time diffusion
stock market models. It is  known that the market completeness is
not a robust property: small random deviations of the coefficients
convert a complete market model into a incomplete one. The paper
shows that market incompleteness is also non-robust: small
deviations can convert an incomplete model into a complete one. More
precisely, it is shown that, for any incomplete market  from a wide
class of models, there exists a complete market model with
arbitrarily close paths of the stock prices and the market
parameters. This leads to a counterintuitive conclusion that the
incomplete markets are indistinguishable from the complete markets
in the terms of the market statistics.
%
\end{abstract}
 \section{Introduction}  The paper studies continuous stock market models and their  statistical
analysis. The possibility of statistical evaluation is investigated
for the market completeness or incompleteness. These concepts are
crucial for the modern mathematical finance. The classical
Black-Scholes market model with a non-random volatility is complete,
meaning that an arbitrarily claim can be replicated with some
self-financing strategies and some initial wealth.   For incomplete
market models, the option replication is not always possible. A
market model with the coefficients that depend on some random factor
being independent from the driving Brownian motion is usually
incomplete (see, e.g, \cite{K}). Typically, the incomplete market
models are used to match the statistical properties of the
historical or implied volatility. Currently, there are many well
developed models for the volatility (see, e.g., \cite{A1}-\cite{A2},
\cite{ab}-\cite{clr}, \cite{eh}-\cite{hw}, \cite{mm}-\cite{z1}). It
is well known that the market completeness is not a robust property:
small random deviations can ruin the completeness and convert a
complete model into a incomplete one.
\par
In the present paper, we address these problems again. We consider a
class of diffusion market models in the setting where the admissible
portfolio strategies  can use historical observations collected
during some time period before the launching time of the replicating
strategy. We found  that the market incompleteness is non-robust
similarly to the market completeness: small deviations can convert
an incomplete market model into a complete one. More precisely,  it
is shown that, for any incomplete market model from a wide class of
models, there exists a complete market model with an arbitrarily
close paths of the stock prices and the market parameters (Theorems
\ref{ThM}-\ref{Th2}). This leads to a conclusion that the incomplete
markets are indistinguishable from the complete markets in the terms
of the market statistics (Corollary \ref{corr1}).
\par
Let us explain why the non-robustness of the incompleteness
established in Theorem \ref{ThM}-\ref{Th2})  leads to a conclusion
that the incomplete markets are indistinguishable from the complete
markets in the terms of the market statistics (Corollary
\ref{corr1}). In theory, for continuous time models, the volatility
can be estimated without error from the historical prices. However,
this would require to know the exact continuous time  path of the
past prices. This is not feasible because the historical prices are
given as time series with rational values. Theorems
\ref{ThM}-\ref{Th2} imply that arbitrarily small rounding and time
discretization errors may lead to different market models with
respect to the completeness and incompleteness. This result is
counterintuitive: there is a common perception that the case of
random volatility leading to the incompleteness can be spotted from
the statistics.
\par
This result does not undermine the importance of the incomplete
market models. These  models reflect the immanent non-predictability
of the real world, in particular, unpredictability of the stock
price volatility.
\par
Theorems \ref{ThM}-\ref{Th2} have rather theoretical than practical
value since they establish some limits for analysis of market
structures based solely on econometrics. These results can be
considered as one more illustration of possibility of  co-existence
of different acceptable models based on the same sets of
observations, in the framework of the concept from
\cite{Madan}-\cite{ME}.
\par
It can be noted that our par follows the general approach to non-robustness of certain market properties
introduced by Guasoniy and  R\'asonyi in \cite{GR}, where non-robustness of arbitrage
opportunities was established.  We
study the incompleteness which is a different market property: the incompleteness  caused by
non-hedgeable randomness of coefficients. The properties considered in this paper and in  \cite{GR}
 neither
exclude nor imply each other. Furthermore, the arbitrage possibility
or completeness are some extreme and rare features.  The arbitrage
possibility  is usually caused by abnormally vanishing volatility or
fast growing appreciation rate; the completeness is caused by the
predictability and the absence of the noise for the volatility. On
the other hand, the incompleteness is rather a typical feature.
Since it is easier to believe that a noise contamination of a model
removes some rare property, the result of the present paper is  more
counterintuitive then the result in \cite{GR}.

\section{The market model}
 We consider  the  so-called diffusion market model,
 where the market dynamic is described by stochastic differential equations
 (see, e.g., \cite{K}\index{Karatzas and   Shreve
(1998) }).  In these equations, the randomness is presented in two
ways: as the white noise being an external input and as the
randomness/uncetainty of the coefficients (market parameters) that
represents the following features: (i) correlations with the past; (ii) non-Markov properties, and
(iii) unpredictability of the future price distributions.

Assume that we are given a  probability space
$(\Omega,\F,\P)$, where $\Omega$ is a set of elementary
events, $\F$ is a complete $\s$-algebra of events  and $\P$ is a
probability measure.

Let $\d>0$ and  $T\in(0,+\infty)$ be given.   Let $w(t)$ be a
Brownian motion defined on $t\in[-\d,T]$ such that $w(-\d)=0$.

Consider continuous time  diffusion model of a securities market
consisting of a risk free bond or bank account with the price
$B(t)$, and
 a risky stock with price $S(t)$, $t\in [-\d,T]$. The prices of the stocks evolve
 as \be \label{S} dS(t)=S(t)\left(a(t)dt+\s(t) dw(t)\right), \quad
 \ee where  $a(t)$ is an
appreciation rate process, $\s(t)$ is a volatility process. The
price of the bond evolves as \baaa \label{B} dB(t)=rB(t)dt, \eaaa
where  $r>0$ is a short rate that is assumed to be constant.
\par
Let $\M$ be the class of random processes $\mu(t)=(a(t),\s(t))$,
$t\in[-\d,T]$, such that the following holds:
\begin{enumerate} \item The processes $a(t)$, $\s(t)$, and $\s(t)^{-1}$ are bounded on $[-\d,T]\times \O$.
\item
$\mu(t)$ is independent from $w(t_2)-w(t_1)$ for all $t,t_1,t_2$
such that $t_2>t_1\ge t\ge -\d$.
 \end{enumerate}

In this paper, we consider market models with $\mu=(a,\s)\in\M$.

Let \index{$\F_t^{w}$ be the filtration generated by the process
$w(t)$, and let} $\F_t^{}$ be the filtration generated by the
process $(w(t),\mu(t))$, $t\ge -\d$. \index{ Let $\F_t^{w}$ be the
filtration generated by the process $w(t\lor 0)$, and let $\F_t^{}$
be the filtration generated by the process $(w(t\lor 0),\mu(t))$,
$t\ge -\d$.}

\par
We assume that $S(-\d)$ and $B(-\d)$ are given non-random variables.
\index{We assume that $S(0)$, and $B(0)$ are given
$\F_0^{}$-measurable random variables. We assume that $S(0)$, and
$B(0)$ are given  non-random variables.} In this case, equation
(\ref{S}) has an unique solution $S(t)$  that is adapted to
$\F_t^{}$, $t\in[-\d,T]$.
\par
By Girsanov Theorem, for any $\mu=(a,\s)\in\M$, there exists a set
${\cal P}_\mu=\{\P_\mu\}$ of  probability measures $\P_\mu$ such
that the process $\ww S(t)=e^{-rt}S(t)$ is a martingale in $t\in[0,T]$
under $\P_\mu$.
 \subsubsection*{Strategies for bond-stock-options market}
We describe below the rules for the operations of the agents on the
market that define the class of admissible strategies that can be
used for replication of contingent claims.
\par
 Let $X(0)>0$
be the initial wealth at time $t=0$, and let $X(t)$ be the wealth
at time $t\in[0,T]$.
\par
We assume that the wealth $X(t)$ at time $t\in[0,T]$ is
\begin{equation}
\label{X} X(t)=\b(t)B(t)+\g(t)S(t).
\end{equation}
Here $\b(t)$ is the quantity of the bond portfolio, $\g(t)$ is the
quantity of the stock  portfolio. The pair of processes $(\b(t),
\g(t))$ describes the state of the bond-stocks securities portfolio
at time $t\in[0,T]$. Each of  these pairs is called a strategy.
\par

The process $\ww X(t)\defi e^{-rt}X(t)$ is called the
discounted wealth, and  the process $\ww S(t)\defi
e^{-rt}S(t)$ is called
 the   discounted stock price, $t>0$.

 A pair $(\b(\cdot),\g(\cdot))$  is said to be an admissible self-financing
strategy if the following holds.
\begin{enumerate}
\item
The processes $\b(t)$ and $\g(t)$ are progressively measurable with
respect to the filtration $\F_t^{}$, $t\in[0,T]$.
\item
There exists  $\P_\mu\in {\cal P}_\mu$ such that
\baaa\E_\mu\int_0^{T}\ww S(t)^2\g(t)^2dt<+\infty, \eaaa where $\E_\mu$
is the expectation with respect to the probability measure $\P_\mu$.
\item  The strategy is  self-financing, meaning that
 $$
dX(t)=\b(t)dB(t)+\g(t)dS(t).$$
\end{enumerate}
For this model, the agents applying admissible self-financing
strategies are not supposed to know the future; the strategies have
to be adapted to the flow of current market information described by
$\F_t^{}$  for $t\in [0,T]$.
\par
The property of self-financing is equivalent to \be\label{wX} d\ww
X(t)=\g(t)d\ww S(t). \ee (See, e.g., \cite{K},\cite{D2002}). It
follows that the process $\g(t)$ alone defines the strategy.
\subsubsection{Market completeness}
\begin{definition} \label{defC} We say that a market model is complete if, for any
$p>0$, any random variable $\xi\in L_{2+p}(\O,\F_T,\P)$ can be
replicated. This means that there exists an $\F_0^{}$-measurable
initial wealth $X(0)$ and an admissible self-financing strategy
$(\b(t),\g(t))$, $t\in[0,T]$, such that the corresponding total
terminal wealth $X(t)$ is such that $X(T)=\xi$ a.s.
\end{definition}
To avoid  technical difficulties, we consider the case where $p>0$ only.

It is well known that the model is complete if the process $\s(t)$
is deterministic.  It is also known that a market model is
incomplete if $\s(\cdot)|_{[0,T]}$ is random and independent from
$w(\cdot)|_{[0,T]}$.   In addition, a model is incomplete if there
is an additional Wiener process $\w w(\cdot)$ that is independent
from $w(\cdot)$ and such that
  the process  $\s(\cdot)|_{[0,T]}$ is not independent from  $\w w(\cdot)|_{[0,T]}$.
\section{The main result}
Let $\M^\bot$ be the set of all $\mu\in\M$ that are independent from
$w(\cdot)$.
\begin{theorem}\label{ThM} For any $\mu\in\M^\bot$, for any $q\ge 1$,
and for any $\e>0$, there exists $\mu_\e\in \M^\bot$ such that the
corresponding market model is complete  and \baa
\E\int_{-\d}^T|\mu_\e(t)-\mu(t)|^qdt+
\E\sup_{t\in[-\d,T]}|S_\e(t)-S(t)|^q<\e. \label{appr}\eaa Here
$S_\e(t)$ is the stock price for the model defined by $\mu_\e$, with
$S_\e(-\d)=S(-\d)$.
\end{theorem}
We denote by $|\mu_\e(t)-\mu(t)|$  the Euclidian norm of the vector.
\begin{corollary}\label{corr1}  The incomplete markets are
indistinguishable from the complete markets in the terms of the
market statistics.
\end{corollary}
\par
{\em Proof of Theorem \ref{ThM}.}  It suffices to consider a market
model with $\mu\in\M$ such that  the market is incomplete.\par
Without a loss of generality, we assume that $a(t)$ and $\s(t)$ are
defined for all $t\in\R$, and that there exists $\d_0>0$ such that
\baaa &a(t)=0,\quad &t\notin[-\d,T+\d],\\  &\s(t)=0,\quad
&t\notin[-\d-\d_0,T+\d_0],\qquad\s(t)=1,\quad
t\in[-\d-\d_0,-\d_0)\cup(T,T+\d_0].\eaaa

 Clearly,
  $a(\cdot,\o)\in L_2(\R)$ and $\s(\cdot,\o)\in L_2(\R)$ for all $\o\in\O$.

Let $\kappa_\e(t)$ be defined as $\kappa_\e(t)=\e^{-1}\kappa_1(t/\e)$,
where $\kappa_1(t)$ is the density for the standard normal
distribution $N(0,1)$.
 Let $\s_\e(t)=\s_\e(t,\o)$ and $a_\e(t)=a_\e(t,\o)$ be the  convolutions
 \baaa
 a_\e(t,\o)=\int_{-\infty}^{\infty}a(s,\o)\kappa_\e(t-s)ds,\quad
\s_\e(t,\o)=\int_{-\infty}^{\infty}\s(s,\o)\kappa_\e(t-s)ds. \eaaa
One may say that $\mu_\e(t)=(a_\e(t),\s_\e(t))$  is the output of a
time invariant smoothing Gaussian filter  representing averaging
with respect to time. It follows that
$\sup_{t,\o}(|a_\e(t,\o)|+|\s_\e(t,\o)|+|\s_\e(t,\o)^{-1}|)$
 is bounded in $\e>0$. Hence $\mu_\e\in\M^\bot$. Note that this filter is not a causal filter
since the output is calculated using the future values of the
process.

For $x\in  L_2(\R)\cup L_1(\R)$, we denote by $X=\digamma x$ the function
$X:\R\to\C$ defined as the Fourier transform of $x$; $$X(\nu)=(\digamma  x)(\nu)=
\int_{-\infty}^{\infty}e^{-i\nu t}x(t)dt,\quad \nu\in\R.$$ If $x\in
L_2(\R)$, then $X$ is defined as an element of $L_2(\R)$.
\par
 Let  \baaa
 \w\kappa_{\e}= \digamma \kappa_\e,\qquad  \w a=\digamma
 a,\qquad
 \w\s=\digamma \s,\quad \w a_{\e}=\digamma a_\e,
 \qquad
 \w\s_{\e}=\digamma \s_\e.\eaaa By the property of convolution,
 \baaa \w a_{\e}(\nu)=\w\kappa_\e(\nu)\w a(\nu),
 \quad
 \w\s_{\e}(\nu)=\w\kappa_\e(\nu)\w \s(\nu),\qquad\nu\in\R.
\eaaa By the properties of the Fourier transform of $\kappa_\e$, for
all $\nu\in\R$,  $\w\kappa_\e(\nu)\to 1$  as $\e\to 0$ a.s. Since
$\mu(t)$ has a finite support on $\R$, we have that $\w a\in
L_2(\R)\cap L_\infty(\R)$ and  $\w \s\in L_2(\R)\cap L_\infty(\R)$
a.s., and the corresponding norms are bounded in $\o$. By Lebesgue's
Dominated Convergence Theorem, \baaa \|\w a_\e(\cdot,\o)- \w
a(\cdot,\o)\|_{L_2(\R)}\to 0,\quad
\|\w\s_\e(\cdot,\o)-\w\s(\cdot,\o)\|_{L_2(\R)}\to
0\quad\hbox{as}\quad \e\to 0\quad \hbox{a.s.}\eaaa It follows that
\baaa \| a_\e(\cdot,\o)- a(\cdot,\o)\|_{L_2(\R)}\to 0,\quad
\|\s_\e(\cdot,\o)-\s(\cdot,\o)\|_{L_2(\R)}\to 0\quad\hbox{as}\quad
\e\to 0\quad \hbox{a.s.}\eaaa It follows  that there exists a
subsequence $\e=\e_i\to 0$ such that $\mu_\e(t,\o)\to \mu(t,\o)$ for
a.e. $t,\o$ as $\e\to 0$. By Lebesgue's Dominated Convergence
Theorem, for any $q\ge 1$, \baaa
\E\int_{-\d}^T|\mu_\e(t)-\mu(t)|^qdt\to 0 \quad\hbox{as}\quad \e\to
0.\eaaa  By Theorem II.8.1 from \cite{Krylov}, it follows that
(\ref{appr}) holds for some $\e_i=\e_i(q,\mu)$.

To complete the proof, it suffices to show that a market model
defined by $\mu_\e(t)=(a_\e(t),\s_\e(t))$ with this $\e=\e_i$ is
complete in the sense of Definition \ref{defC}.
\par
Let $\A_t^\e$ be the filtration generated by the process
$\mu_\e(t)$, $t\ge -\d$.
By Proposition 3 from \cite{D2010a}, the process $\mu_\e(t,\o)$ is
weakly predictable for any $\o\in\O$ meaning that, for any $f\in
L_\infty(0,T)$, the integrals $\int_0^Ta_\e(t,\o)f(t)dt$ and
$\int_0^T\s_\e(t,\o)f(t)dt$ can be found   with an arbitrarily small
error using the values $\{\mu_\e(\tau,\o)\}_{\tau\le 0}$. Moreover,
by Proposition 1 \cite{D2010a},   the processes $a_\e(t,\o)$ and
$\s_\e(t,\o)$ are analytic functions in $t$ for all $\o\in\O$. It
follows that $\mu_\e(t)$ is a $\A^{\e}_0$-measurable random vector
for any $t\in[0,T]$.
\par
Let $\F^\e_t$ be the filtration generated  by the process $(w(t),\mu_\e(t))$, $t\ge -\d$. In other words, this
filtration is generated
 by the
observations $\{w(s),\ \mu_\e(s),\ -\d<s<t\}$.   We have established
that the analytic properties of $\mu_\e(t)$ imply that the same
filtration is generated  by the process $(w(t),\mu_\e(t\land 0))$,
$t\ge -\d$, i.e., this filtration is generated by the observations
$\{w(s),\ \mu_\e(s\land 0),\ s\le t\}$.
\par
Let $\t_\e(t)=\s_\e(t)^{-1}(a_\e(t)-r)$ and  let
$w_\e(t)=\int_0^t\t_\e(s)ds +w(t)-w(0)$. Let a probability measure
$\P_\e$ be defined such that \baaa
\frac{d\P_\e}{d\P}=\exp\left(-\frac{1}{2}\int_0^T\t_\e(t)^2dt-\int_0^T\t_\e(t)dw(t)\right).
\eaaa Let $\E_\e$ be the corresponding expectation. By Girsanov
Theorem applied on the conditional probability space given
$\F^{\e}_0$,  the process $w_\e(t)$ is a Wiener process
conditionally given $\F^{\e}_0$ under the conditional probability
measure $\P_\e(\,\cdot\,|\F^{\e}_0)$. \index{Let $\P_\e$ be the
probability measure such that the corresponding expectation
$\E_\e\zeta$ of a random variable $\zeta$ is
$\E_\e\zeta=\E\E_\e\{\zeta|\F^{\e}_0\}$, where
$\E_\e\{\,\cdot\,|\F^{\e}_0\}$ is the expectation for the
conditional measure $\P_\e(\,\cdot\,|\F^{\e}_0)$.}
\par
By the assumptions on $\xi$, $\E\xi^{2+p}<+\infty$ for some $p>0$.
Hence \baaa \E_\e\xi^2<+\infty,\quad
\E_\e\{\xi^2|\F^{\e}_0\}<+\infty\quad\hbox{a.s..}\eaaa By the
Martingale Representation  Theorem applied on the conditional
probability space given $\F^{\e}_0$, there exists a process
$g_\e=g_\e(t,\o)$ such that $g_\e(t)$ is adapted to $\F^{\e}_t$ and
\baaa \E_\e\int_0^Tg_\e(t)^2dt<+\infty,\quad
\E_\e\left\{\int_0^Tg_\e(t)^2dt\Bigl|\F^{\e}_0\right\}<+\infty\quad
\hbox{a.s.}, \eaaa \index{ $g_\e\in
L_2(\O,\F,\P_\e(\,\cdot\,|\F^{\e}_0),L_2(0,T))$ a.s., $g\in
L_2(\O,\F,\P_\e,L_2(0,T))$} and \baaa
e^{-rT}\xi=e^{-rT}\E_\e\{\xi|\F^{\e}_0\}+\int_0^Tg_\e(t)dw_\e(t).
\eaaa (See, e.g., Theorem 4.2.4 in \cite{LL}, p.67). It follows that
\baaa
e^{-rT}\xi=e^{-rT}\E_\e\{\xi|\F^{\e}_0\}+\int_0^T\g_\e(t)dS_\e(t),
\eaaa where $\g_\e(t)=g_\e(t)\s_\e(t)^{-1}\ww S_\e(t)^{-1}$, where
$\ww S_\e(t)=e^{-rt}S_\e(t)$. By (\ref{wX}), it follows that the
self-financing strategy with the initial wealth
$X_\e(0)=e^{-rT}\E_\e\{\xi|\F^{\e}_0\}$ and with the quantity of
shares $\g_\e(t)$ is such that the terminal discounted wealth $\ww
X_\e(T)$ is $e^{-rT}\xi$. Hence the terminal wealth for this
strategy is $X_\e(T)=\xi$. This completes the proof.
 $\Box$
\begin{remark}
{\rm In our setting, it is essential that the initial wealth
$X_\e(0)$ for the replicating strategy  is $\F^{\e}_0$-measurable,
where $\F^{\e}_t$ is the filtration describing the information flow
for $t\ge -\d$, and that $\F^{\e}_0$ is a non-trivial. The
information about the history before $t=0$ is used for the
predicting $\mu_\e(t)$ for $t\in[0,T]$. This is what makes the
approximating market model complete. }\end{remark}
\section{An economic interpretation} Theorem \ref{ThM} implies that
the selection of a incomplete model cannot be based solely on the
market statistics. This does not undermine a practical use of
incomplete market models. Selecting these models, we admit the
immanent non-predictability of the real world. For instance, we
would rather accept a model with the possibility of the
unpredictable jumps for the volatility than a model where these
jumps can be predicted, even if the statistical data supports both
models equally.

Let us discuss the consequences of co-existing of statistically
indistinguishable complete and incomplete markets models.

In the proof of Theorem \ref{ThM}, the process $\mu(t)$ is
approximated by an analytic function $\mu_\e(t)$ that is  used to
set  a new alterative model. For the new model, the future values
$\mu_\e(t)$ are uniquely defined by their values on the time
interval $[\d,0]$. However, since the new and the old models produce
arbitrarily close sets of prices, an observer cannot tell apart
these models with certainty, i.e., she cannot tell which model
generates the observed data. Effectively, the process $\mu_\e(t)$ in
the new model is not observable for an observer from the old model.

It can be further illustrated  as the following.  Assume
that an option trader has collected the marked data $t\in [-\d,0]$ with
the purpose to test the following hypotheses  $H_0$ and $H_A$ about
the stock price evolution:
\begin{itemize}\item $H_0$: the values $\mu(t)|_{t\in[0,T]}$ are not
$\F_0$-measurable for $t\in[0,T]$ (i.e., the market is incomplete).
\item $H_A$: the values  $\mu(t)|_{t\in[0,T]}$ are
$\F_0$-measurable for $t\in[0,T]$ (i.e., the market is complete).
\end{itemize}
\par
It can be noted that we can replace the hypothesis $H_0$ by a
hypothesis assuming a particular stochastic volatility model, such
as a Markov chain model,  Heston  model, etc.

The trader has  to calculate at time $t=0$  the price of an option
expiring at time $T$; different hypothesis lead to different prices.
According to Theorem \ref{ThM}, it is impossible to reject $H_A$
hypothesis based solely on market statistics collected during the
time period $[-\d,0]$.  \index{If this hypothesis was rejected
anyway, then, according to Theorem \ref{ThM}, one can say that
certain beliefs were involved besides the market statistics, such as
beliefs that some unexpected jumps of the parameters may occur.
Therefore, it can be concluded that the selection of the stock price
model for the option pricing is never based solely on the market
statistics. }

Due to rounding errors, the statistical indistinguishability  leading to this conclusion
cannot be fixed via the sample increasing since the
statistics for the incomplete market models can be arbitrarily close
to the statistics of  the alternative complete models. \index{Of course, separation of the hypotheses  about the
distributions or evolution laws for processes  by statistical
methods is never possible with 100\% confidence. However, the
probability of the error is usually decreasing  for large samples.}

It can also noted that, unfortunately, the predictability of
$\mu_\e$ cannot be used for option pricing under the "natural"
hypothesis $H_0$. The stock prices and market parameters  under
these two hypotheses are pathwise close; however, their properties
are quite different with respect to the predicability. The process
$\mu_\e(t)$ is an output of a non-causal smoothing filters, and its
calculation would require the future values of $\mu(t)$ that are
unavailable in practice.

\section{A more general setting} In the previous section, we
considered $\mu\in\M^\bot$, i.e., $\mu$ was assumed to be
independent from the driving Wiener process. In fact, this
assumption was rather technical; analogs of Theorem \ref{ThM} can be
obtained for more general models where $\mu(\cdot)$  can depend on
$w(\cdot)$ or $S(\cdot)$. Let us give an example.

  Let $y(t)=y( t,\o)$ be a bounded random process with the
values at $\R^N$, $t\in[-\d,T]$, such that $y$ is independent from $w$.

Let $R(t)=\log S(t)$. Let   $\d_0>0$ be given, and let $\ww\M$ be the class of all $\mu\in\M$
allowing a closed-loop representation \baa
\mu(t)=(a(t),\s(t))^\top=M(y(t),R(t),\oo w(\cdot),t).\label{M} \eaa In
(\ref{M}),   $M$ is a measurable bounded  function $M:\R^N\times \R\times
C(-\d-\d_0,T+\d_0)\to \R^2$,  $\oo w(t)=w((-\d\lor t)\land t)$. We assume that $M$ is such that the following holds:
\begin{enumerate}
\item
$M(y,\rho,\xi,t)$ is continuous in $y\in\R^N$ uniformly in
$(\rho,\xi,t)\in \R\times C(-\d-\d_0,T+\d_0)\times [-\d-\d_0,T+\d_0]$.
\item $M(y,\rho,\xi,t)$ is Lipschitz   in $\rho\in\R$ uniformly in $(y,\xi,t)\in
\R^N\times C(-\d-\d_0,T+\d_0)\times [-\d-\d_0,T+\d_0]$.
\end{enumerate}
The choice of $y$ and $M$ defines  $\mu\in\M$.

Consider equation for the process $R(t)=\log S(t)$ \baa
dR(t)=a(t)dt-\frac{\s(t)^2}{2}dt+\s(t)dw(t),\qquad R(0)=\log S(0).
\label{RS}\eaa
For any $\mu\in\ww\M$, the assumptions on $M$ ensure existence of an
unique solution of equation (\ref{RS}).
This implies solvability of (\ref{S}) with $S(t)=e^{R(t)}$.

We introduce the market model,  admissible strategies, and the definition of completeness
such as defined above but with the filtration $\F_t$ redefined as the filtration generated by the process $(w(t),y(t))$.
\begin{theorem}\label{Th2} For any $\mu\in\ww\M$, for any $q\ge 1$,
and for any $\e>0$, there exists $\mu_\e\in \ww\M$ such that the
corresponding market model is complete  and \baaa
\E\int_{-\d}^T|\mu_\e(t)-\mu(t)|^qdt+ \E\sup_{t\in[-\d,T]}|\log
S_\e(t)-\log S(t)|^q<\e. \label{apprlog}\eaaa Here $S_\e(t)$ is the
stock price for the model defined by $\mu_\e$ such that
$S_\e(0)=S(0)$.
\end{theorem}
\par
Note that Theorem \ref{Th2} is a generalization of Theorem \ref{ThM}
since $\M^\bot\in\ww\M$; the models in Theorem \ref{ThM} belong to
the class $\ww\M$ with $N=2$ and with $y(t)=\mu(t)=(a(t),\s(t))^\top$.
\par
{\em Proof of Theorem \ref{Th2}.} Without a loss of generality, we assume that
$y(t,\o)=0$ for
$t\notin[-\d-\d_0,T+\d_0]$ for all $\o$.  We use approximating models with
$\mu_\e(t)=M(y_\e(t),R_\e(t),w(\cdot),t)$, where \baaa
y_\e(t)=\int_{-\infty}^{\infty}y(s)\kappa_\e(t-s)ds\eaaa is the
output of the Gaussian smoothing filter, and where $R_\e(t)=\log
S_\e(t)$ is the solution of the corresponding equation (\ref{RS})
such that $R_\e(-\d)=\log S(-\d)$.  By the definitions,
$\mu_\e\in\ww\M$. The market model for $\mu\in\ww\M$ is arbitrage
free and complete if $y(t)$ is a bounded deterministic process for
$t\in[0,T]$. The rest of the proof follows the proof of Theorem
\ref{ThM}. The assumption (iii) on $M$ ensures applicability of
Theorem II.8.1 from \cite{Krylov} to equations (\ref{RS}) with
$\mu=\mu_\e$. $\Box$
\section{Concluding remarks on forecasting and future development}\label{secCon}
 We outline below some possible modifications and future developments.\index{,  with respect to forecasting technique for the market parameters
 used in  the proofs. }\begin{enumerate}
\item
Theorems \ref{ThM}-\ref{Th2} allow other modifications. For
instance, the statement of Theorem \ref{ThM} holds with $\M^\bot$
replaced by a class $\hat\M$ of all $\mu\in\ww\M$ such that (\ref{M}) holds with $M$ such that
$sM(y,\log s,\xi,t)$ is Lipschitz   in $s\in (0,+\infty)$ uniformly in
$(y,\xi,t)\in R^N\times C(-\d-\d_0,T\d_0)\times [-\d-\d_0,T+d_0]$.
\item
The predictability used in the proof of Theorem \ref{ThM} can be
ensured by many different non-causal time invariant smoothing
filters. Instead of a Gaussian filter, we can use an ideal low pass
filter or a filter with the exponential rate  of energy on higher
frequencies $e^{-|\nu|T}$. The output of a process transferred with
these smoothing filters is a process that, at time $t=0$,  can be
predicted   on time interval $[0,T]$ (see \cite{D2010a}).
\item  Currently, it is unknown  if a Gaussian
filter can be approximated  by causal smoothing filters.  It is
known that the approximation  by causal smoothing filters is
impossible for the ideal low pass
 filters; the distance of the set of the ideal low-pass  filters from the set
of all causal filters is positive \cite{rema}.  On the other hand,
it is known that a filter with the exponential energy decay  allows
arbitrarily close approximation by causal filters \cite{D2012a}.
This could lead to application of filters with the exponential
energy decay on higher frequencies  for forecasting of market
parameters and approximation of $\mu_\e$. It could be interesting to
explore this opportunity.

\item It could be interesting  to extend the approach of this paper on
discrete market models. For this, discrete time predictability criterions from
\cite{D2012b}-\cite{D2012c} could be used.
\end{enumerate}
\subsection*{Acknowledgment} \iindex{ This work  was supported by ARC grant of Australia DP120100928 to the
author.}

\end{document}